\documentclass[letterpaper, 10 pt, conference]{ieeeconf}  % Comment this line out if you need a4paper

\IEEEoverridecommandlockouts
\usepackage{graphicx,times,amsmath,amssymb,mathrsfs,cite,stfloats,booktabs,url,multirow, subfigure}
\usepackage[lined,ruled,commentsnumbered]{algorithm2e}

\title{\LARGE \bf Protecting Multiple Types of Privacy Simultaneously \\ in EEG-based Brain-Computer Interfaces}

\author{Lubin~Meng, Xue~Jiang, Tianwang~Jia, Dongrui~Wu
\thanks{Key Laboratory of Image Processing and Intelligent Control, School of Artificial Intelligence and Automation, Huazhong University of Science and Technology, Wuhan, China.}
\thanks{Corresponding author: D.~Wu (e-mail: drwu@hust.edu.cn)}}

\begin{document}

\maketitle
\thispagestyle{empty}
\pagestyle{empty}

\begin{abstract}

A brain-computer interface (BCI) enables direct communication between the brain and an external device. Electroencephalogram (EEG) is the preferred input signal in non-invasive BCIs, due to its convenience and low cost. EEG-based BCIs have been successfully used in many applications, such as neurological rehabilitation, text input, games, and so on. However, EEG signals inherently carry rich personal information, necessitating privacy protection. This paper demonstrates that multiple types of private information (user identity, gender, and BCI-experience) can be easily inferred from EEG data, imposing a serious privacy threat to BCIs. To address this issue, we design perturbations to convert the original EEG data into privacy-protected EEG data, which conceal the private information while maintaining the primary BCI task performance. Experimental results demonstrated that the privacy-protected EEG data can significantly reduce the classification accuracy of user identity, gender and BCI-experience, but almost do not affect at all the classification accuracy of the primary BCI task, enabling user privacy protection in EEG-based BCIs.

\end{abstract}

\section{Introduction}

A brain-computer interface (BCI) enables the user to interact with or control an external device (computer, wheelchair, robot, etc.) using brain signals. The electroencephalogram (EEG)~\cite{Lance2012}, which captures the brain's electrical activities from the scalp, is the most popular input of BCIs, due to its convenience and low cost. EEG-based BCIs have found successful applications in neurological rehabilitation~\cite{Daly2008}, emotion recognition~\cite{drwuPIEEE2023}, robotic device control~\cite{Edelman2019}, and so on.

Machine learning has achieved great successes in BCIs, which recognizes complex patterns~\cite{drwuSF2018} in EEG signals and builds high-performance classification/regression models~\cite{drwuMITLBCI2022}. Typically, a substantial amount of EEG data is required to train an accurate machine learning model. However, EEG data not only capture task-specific information but also include significant personal and private details~\cite{drwuTCSS2023}. For instance, Martinovic \emph{et al.}~\cite{Martinovic2012} demonstrated that EEG signals have the potential to disclose various private details, such as credit cards, PIN numbers, known people, and residential addresses. Choi \emph{et al.}~\cite{Choi2018} found that user identity can be inferred from resting-state EEG signals with an accuracy of 88.4\%. Meng \emph{et al.}~\cite{Meng2023a} further showed that user identity can be easily inferred from EEG signals across different BCI tasks. Kaushik \emph{et al.}~\cite{Kaushik2019} revealed that user gender and age could be predicted by analyzing resting-state EEG recordings.

In response to growing privacy concerns, numerous laws have been enacted worldwide, such as the General Data Protection Regulation of the European Union and the Personal Information Protection Law of China, aimed at enforcing stringent user privacy protections. Consequently, multiple privacy-protection techniques have been developed for EEG-based BCIs. They can be categorized into two groups. The first is cryptographic, including secure multiparty computation, homomorphic encryption, and secure processors. Agarwal \emph{et al.}~\cite{drwuPrivacy2019} introduced cryptographic protocols based on secure multiparty computation to safeguard privacy in EEG-based driver drowsiness estimation. The second category is privacy-preserving machine learning, enabling machine learning without directly accessing raw EEG data or model parameters. Typical approaches include federated learning and source-free transfer learning. Xia \emph{et al.}~\cite{Xia2022} proposed augmentation-based source-free adaptation, which enables privacy-preserving transfer learning without accessing the source EEG data and/or model parameters. Zhang and Wu~\cite{drwuLSFT2022} used lightweight source-free transfer to address a similar issue. Zhang \emph{et al.}~\cite{drwuMSDT2022} further proposed unsupervised multi-source decentralized privacy-preserving transfer.

Cryptographic and privacy-preserving machine learning safeguard privacy by restricting EEG data sharing. However, this significantly constrains the  data availability in many scenarios. For example, the model generated by privacy-preserving machine learning may not be optimal, and without access to the EEG data, users cannot further improve the models or design better algorithms. On the other hand, if EEG datasets could be publicly shared, it may accelerate research and discovery, particularly in the development of large models which typically require extensive data for training. To balance privacy protection and data accessibility, perturbation was widely used for privacy protection, which adds noise to or transforms the original data to conceal private information while maintaining the downstream task information. Meng \emph{et al.}~\cite{Meng2023a} deliberately designed two perturbations to convert the original EEG data into identity-unlearnable EEG data, which can be used to protect the user identity privacy while maintaining the primary BCI task performance. However, the generated perturbations can only protect a single type of private information (identity) in the EEG data.

Broader access to EEG data necessitates more comprehensive privacy safeguards. This paper further demonstrates that in addition to the user identity information, other private information (e.g., gender, BCI-experience) can also be inferred from EEG data. To protect multiple types of privacy simultaneously, we design the perturbation to convert the original EEG data into privacy-protected EEG data. Specifically, we first generate tiny perturbations highly correlated with each type of private information separately, which can conceal the private information without affecting BCI task related information. Then, the superposition of perturbations corresponding to different types of private information is used to convert the original EEG data into privacy-protected EEG data.

In summary, we make the following contributions:
\begin{enumerate}
	\item We expose a serious privacy problem in EEG-based BCIs, i.e., multiple types of private information (user identity, gender, and BCI-experience) can be inferred from EEG signals.
	\item We propose an approach to convert the original EEG data into privacy-protected EEG data, which can protect multiple types of private information simultaneously while maintaining the primary BCI task performance.
\end{enumerate}

\section{Method}

This section introduces the details of protecting multiple types of private information in EEG data simultaneously.

\subsection{Problem Statement}

Given an EEG dataset $\mathcal{D}=\{(\mathbf{x}_i, y_i, \mathcal{S}_i)\}_{i=1}^{N}$, where $\mathbf{x}_i \in \mathcal{X} \subset \mathbb{R}^{c\times t}$ is the $i$-th EEG trial with $c$ channels and $t$ time domain samples, $y_i \in \mathcal{Y}=\{1,...,K\}$ the task label (e.g., target or non-target in event related potential classification), $\mathcal{S}_i=\{p_{i,m}\}_{m=1}^{M}$ the set of privacy label that can be inferred from the EEG trial $\mathbf{x}_i$, and $p_{i,m}\in \mathcal{P}_m=\{1,...,P_m\}$ a specific privacy label (e.g., male and female in gender).

Typically, a Task-Classifier $F$ can be trained on $\mathcal{D}$ to learn the mapping from the EEG input space to the task label space, i.e., $F:\mathcal{X}\rightarrow \mathcal{Y}$. However, EEG data not only record task-related information but may also contain rich personal privacy. So, a Privacy-Classifier $G_m$ can be constructed from EEG data to mine the private information, i.e., $G_m:\mathcal{X}\rightarrow \mathcal{P}_m$ maps the EEG input space into a specific privacy space.

This paper aims to generate perturbations for the original EEG data to protect multiple types of private information simultaneously, while preserving the task-related information to ensure the normal usage. Specifically, we generate a perturbed EEG dataset $\mathcal{D}'=\{(\mathbf{x}'_i, y_i, \mathcal{S}_i)\}_{i=1}^{N}$, where $\mathbf{x}'=\mathbf{x}+\boldsymbol{\delta}$ and $\boldsymbol{\delta}$ is a deliberately designed perturbation. For any privacy $\mathcal{P}_m$, it is difficult to train a Privacy-Classifier $G_m$ from the perturbed dataset, making the private information in EEG data unlearnable. However, this perturbed dataset can still be used to train a good Task-Classifier $F$ as the original unperturbed dataset $\mathcal{D}$.

\subsection{Perturbation Generation}

To prevent Privacy-Classifiers from learning private information in the EEG data, we design perturbations highly correlated with the private information, which can mislead the Privacy-Classifier to learn the perturbation pattern rather than the true privacy pattern. Additionally, due to the distribution differences between the privacy feature space and the task feature space, the perturbations generated for privacy protection may have little impact on the primary BCI task, which will be verified in Sections~\ref{sect:G} and \ref{sect:H}.

Specifically, we first train a Privacy-Classifier $G'_m$ for each privacy type $\mathcal{P}_m$ to evaluate the influence of perturbations on the privacy-related information, which may be different from $G_m$. The loss function for $G'_m$ is
 \begin{align}
 	\min_{\boldsymbol{\theta}_{G'_m}}\mathbb{E}_{(\mathbf{x},p_m)\sim \mathcal{D}}\ell_{\mathrm{CE}}(G'_m(\mathbf{x}), p_m), \label{eq:model}
 \end{align}
where $\boldsymbol{\theta}_{G'_m}$ is the parameter set of $G'_m$, and $\ell_{\mathrm{CE}}$ the cross-entropy loss.

Then, for each privacy $\mathcal{P}_m$, we generate a class-wise perturbation $\boldsymbol{\delta}_m=[\boldsymbol{\delta}_{m,1},...,\boldsymbol{\delta}_{m,P_m}]$ by minimizing following loss function:
 \begin{align}	\min_{\boldsymbol{\delta}_m}\mathbb{E}_{(\mathbf{x},p_m)}\left[\ell_{\mathrm{CE}}(G'_m(\mathbf{x}+\boldsymbol{\delta}_{m,p_m}),p_m)+\alpha\|\boldsymbol{\delta}_{m,p_m}\|_2\right], \label{eq:perturbation}
 \end{align}
where $\boldsymbol{\delta}_{m,p_m}$ is the perturbation for class $p_m$ of privacy type $\mathcal{P}_m$, and $\alpha$ is a trade-off parameter. The first term enhances the correlation between the perturbation $\boldsymbol{\delta}_{m,p_m}$ and the privacy class $p_m$, and the second term constrains the perturbation amplitude.

After generating perturbation for each privacy type, we can add them to each EEG trial $\mathbf{x}$ in $\mathcal{D}$ to protect multiple different types of privacy simultaneously:
 \begin{align}
 	\mathbf{x}'=\mathbf{x}+\sum_{m=1}^M \boldsymbol{\delta}_{m,p_m}. \label{eq:addtion}
 \end{align}

Finally, the EEG dataset with privacy protection is $\mathcal{D}'=\{(\mathbf{x}'_i, y_i, \mathcal{S}_i)\}_{i=1}^{N}$. Since the perturbation for each privacy type is very small, the aggregated perturbation is still small enough to have little impact on the primary task performance.

Algorithm~\ref{alg:alg} gives the pseudo-code of privacy-protected EEG dataset generation.

\begin{algorithm}[htpb]
\KwIn{$\mathcal{D}=\{(\mathbf{x}_i, y_i, \mathcal{S}_i)\}_{i=1}^N$, the original EEG dataset\;
\hspace*{11mm}$\{G'_m\}_{m=1}^M$, the set of Privacy-Classifiers\;
\hspace*{11mm}$T_{\mathrm{m}}$, the maximum number of model training epochs\;
\hspace*{11mm}$T_{\mathrm{p}}$, the maximum number of perturbation optimization epochs\;}
\KwOut{A privacy-protected EEG dataset $\mathcal{D}'=\{(\mathbf{x}'_i,y_i,\mathcal{S}_i)\}_{i=1}^N$.}

\tcp{Perturbation generation for each single privacy type}
\For{$m=1,...,M$}{
Initialize $\boldsymbol{\delta}_m \leftarrow \mathcal{N}(0,0.001)$\;	
\For{$t=1,...,T_{\mathrm{m}}$}{
Update $G'_m$ by (\ref{eq:model}) on $\mathcal{D}$ \;
}	

\For{$t=1,...,T_{\mathrm{p}}$}{
Update $\boldsymbol{\delta}_m$ by (\ref{eq:perturbation}) on $\mathcal{D}$ \;
}
}
\tcp{Perturbation generation for all privacy types}
\For{$i=1,...,N$}{
Calculate the perturbed EEG trial $\mathbf{x}'_i$ by (\ref{eq:addtion})\;
}

\textbf{Return} $\mathcal{D}'=\{(\mathbf{x}'_i,y_i,\mathcal{S}_i)\}_{i=1}^N$.
\caption{Privacy-protected EEG dataset generation.} \label{alg:alg}
\end{algorithm}

\section{Experiments}

This section introduces the experimental settings and results for validating the performance of the privacy-protected EEG data.

\subsection{Datasets}

The publicly available EEG dataset introduced by Lee \emph{et al.}~\cite{Lee2019} was used in our experiments. It was collected from 54 healthy subjects, among which 16 were experienced BCI users. Each subject performed a sequence of three tasks: a 36-symbol event-related potential (ERP) ~\cite{Hueebner2018} task, a binary-class motor imagery (MI)~\cite{drwuMITLBCI2022} task, and a four-target steady-state visually evoked potential (SSVEP)~\cite{Chen2015b} task. During each task, 62-channel EEG data were recorded with a sampling rate of 1,000Hz. The entire experimental procedure was repeated twice, so the dataset includes two sessions. The details of each task are as follows:
\begin{enumerate}
	\item ERP: In the ERP task, the subjects were instructed to focus on the target symbol which was flashed randomly to elicit a P300 response. The goal was to classify whether the target symbol flashed or not. The collected EEG data for the ERP task contained 4,140 trials.
	\item MI: In the MI task, the subjects performed the imagery task of grasping with the corresponding hand when the left or right arrow cue appeared. The collected EEG data for the MI task contained 200 trials with balanced left and right hand imagery tasks.
	\item SSVEP: In the SSVEP task,  there were four blocks with different flickering frequencies presented in four positions on a monitor, and the subjects were asked to gaze at the target block. The goal was to classify which block the subject gazed at. The collected EEG data for the SSVEP task contained 200 trials, with 50 trials per block.
\end{enumerate}

For preprocessing, we downsampled EEG trials to 128Hz and applied a [1,40]/[4,40]/[4,64]Hz band-pass filter for ERP/MI/SSVEP, respectively. Next, we extracted EEG trials between [0,2]s after each task stimuli and standardized the EEG trials separately for each subject, task, and session using $z$-score normalization. To balance the data volume for each task, we downsampled each session of the ERP task to 200 trials.

Before the experiments, the subjects filled out a questionnaire to record their personal information and to check their physical and mental condition, which are all personal privacies. The subjects' identity, gender and BCI-experience (whether they had prior BCI experimental experience or not) from the questionnaire were used in our experiments. We first validated that these three types of private information can be inferred from the original EEG data, and then designed perturbations to protect them.

\subsection{Models}

We used the following three convolutional neural networks (CNN) as Privacy-Classifiers:
\begin{enumerate}
 	\item EEGNet~\cite{Lawhern2018}: EEGNet is a compact CNN architecture tailored for EEG-based BCIs. It comprises two convolutional blocks, utilizing depthwise and separable convolutions instead of traditional convolutions, to reduce the number of model parameters.
 	\item DeepCNN~\cite{Schirrmeister2017}: DeepCNN is composed of four convolutional blocks. The initial block is tailored for processing EEG data, while the subsequent three blocks follow a standard convolutional design.
 	\item ShallowCNN~\cite{Schirrmeister2017}: ShallowCNN is a more straightforward variant of DeepCNN, drawing from filter bank common spatial patterns. It features a single convolutional block with a larger kernel and employs a different pooling strategy compared to DeepCNN
\end{enumerate}

\subsection{Performance Measure}

Given class-imbalance in gender and BCI-experience classification, balanced classification accuracy (BCA) was used to evaluate the Privacy-Classifier's performance:
 \begin{align}
 	\mathrm{BCA}_m = \frac{1}{P_m}\sum_{p=1}^{P_m}\frac{1}{N_{p_m}}\sum_{i=1}^{N_{p_m}}\mathbb{I}(G_m(\mathbf{x}_i)=p_m),
 \end{align}
where $P_m$ is the number of classes for privacy type $\mathcal{P}_m$, $N_{p_m}$ the number of test samples in class $p_m$, and $\mathbb{I}(\cdot)$ an indicator function.

\subsection{Experimental Settings}

We performed leave-one-session-out cross-validation, i.e., trials from three tasks in one session were mixed as the training set, and the other session was used as the test set. The average BCA from two sessions were computed. The entire cross-validation process was repeated 5 times, and the average results are reported.

\subsection{Identity, Gender, and BCI-experience Privacy Mining}

The original EEG signals contain rich personal private information. To demonstrate that, Privacy-Classifiers were trained on the unperturbed EEG dataset to classifier user identity, gender, and BCI-experience. The test performance of Privacy-Classifiers are shown in the `Unperturbed EEG' panel of Table~\ref{tab:main_result}. Regardless of which CNN model (EEGNet, DeepCNN, or ShallowCNN) was used as the Privacy-Classifier, the BCAs for the three types of private information were much higher than random guess. These results confirmed that rich personal private information can be inferred from EEG data across different BCI tasks, highlighting the necessity of privacy protection in EEG-based BCIs.

\subsection{Identity, Gender, and BCI-experience Privacy Protection}

To safeguard the private information contained in EEG data, we converted the original EEG data into privacy-protected EEG data, making it difficult for machine learning models to learn the private information.

Specifically, we generated privacy-protected EEG data as described in Algorithm~\ref{alg:alg}. The test performance of Privacy-Classifiers trained on privacy-protected EEG data are shown in the `Perturbed EEG' panel of Table~\ref{tab:main_result}. For each privacy type (identity, gender, and BCI-experience) classification, the BCAs of the Privacy-Classifiers trained on the perturbed EEG data were significantly lower than their counterparts on the original EEG data. Especially, for gender and BCI-experience classification, the BCAs after perturbations were close to random guess, demonstrating that little private information can be learned from the privacy-protected EEG data.

Notice that although the privacy-protected EEG data were generated by EEGNet, they remained effective for DeepCNN and ShallowCNN, indicating good generalization of privacy-protected EEG data.

\begin{table}[!t] \centering \setlength{\tabcolsep}{0.9mm}
\caption{BCAs (\%) of Privacy-Classifiers on the original EEG data and the perturbed EEG data. }
\begin{tabular}{c|c|c|c|c|c}
\toprule
Privacy & Number of& Privacy & Original & Perturbed  & Average \\
 	Type& Classes & Classifier & EEG &EEG & Reduction \\ \midrule
 \multirow{3}{*}{Identity} & \multirow{3}{*}{54} & EEGNet & 25.55  & 4.52  & 20.03   \\
    & & DeepCNN & 34.98  & 3.26  & 31.72    \\
    & & ShallowCNN & 49.13  & 9.76  & 39.37  \\ \midrule
\multirow{3}{*}{Gender} & \multirow{3}{*}{2} & EEGNet & 81.01  & 50.04  & 30.97  \\
    & & DeepCNN & 78.31  & 51.03  & 27.28 \\
    & & ShallowCNN & 82.31  & 51.24  & 31.07  \\  \midrule
    BCI- & \multirow{3}{*}{2} & EEGNet & 67.44  & 50.06   & 17.38  \\
    Experience & & DeepCNN & 70.17  & 51.09   & 19.08   \\
    & & ShallowCNN & 72.34  & 50.33  & 22.01   \\\midrule
\multicolumn{3}{c|}{Average}  & 62.25 & 35.70 & 26.55  \\   \bottomrule
\end{tabular}
\label{tab:main_result}
\end{table}

\subsection{BCI Task Performance} \label{sect:G}

In addition to privacy protection, the perturbations should also minimize the impact on the primary BCI tasks, i.e., the performance of Task-Classifiers trained on privacy-protected EEG data should be similar to their counterparts on the original unperturbed EEG data.

Table~\ref{tab:task_result} shows the test performance of the three CNN models and a traditional model [i.e., xDAWN~\cite{Rivet2009} filtering and Logistics Regression (LR) classification for  ERP, common spatial pattern (CSP)~\cite{Ramoser2000} filtering and LR classification for MI, and canonical correlation analysis (CCA)~\cite{Lin2007} for SSVEP] as Task-Classifier trained on the privacy-protected EEG data and the original unperturbed EEG data. The BCAs after applying perturbations were very close to their counterparts on the original unperturbed EEG, suggesting that privacy-protected EEG had almost no negative impact on the performance of primary BCI tasks.

\begin{table}[!t] \centering \setlength{\tabcolsep}{1.0mm}
\caption{BCAs (\%) of Task-Classifiers on the original EEG data and perturbed EEG data. }
\begin{tabular}{c|c|c|c|c|c}
\toprule
BCI& Number of & Task & Original & Perturbed  & Average \\
Task 	& Classes & Classifier & EEG &EEG & Reduction \\ \midrule
\multirow{4}{*}{ERP} & \multirow{4}{*}{2} & EEGNet & 82.44  & 82.15  & 0.29   \\
    & & DeepCNN & 82.62  & 82.53  & 0.09    \\
    & & ShallowCNN & 79.43  & 79.04  & 0.39  \\
    & & xDAWN+LR & 78.36 & 76.32 & 2.04  \\ \midrule
\multirow{4}{*}{MI} & \multirow{4}{*}{2} & EEGNet & 74.35  & 74.37  & -0.02  \\
    & & DeepCNN & 77.59  & 77.66  & -0.07 \\
    & & ShallowCNN & 75.67  & 75.57  & 0.10  \\
    & & CSP+LR & 63.58 & 64.08 & -0.50  \\ \midrule
\multirow{4}{*}{SSVEP} & \multirow{4}{*}{4} & EEGNet & 95.51  & 95.40   & 0.11  \\
    & & DeepCNN & 95.95  & 95.82  & 0.13   \\
    & & ShallowCNN & 93.59  & 92.57  & 1.02   \\
    & & CCA & 90.30 & 90.30 & 0.00 \\ \midrule
\multicolumn{3}{c|}{Average}  & 82.45 & 82.15 & 0.30  \\   \bottomrule
\end{tabular}
\label{tab:task_result}
\end{table}

\subsection{Characteristics of the Privacy-Protected EEG Data} \label{sect:H}

We further show that the characteristics of the EEG data before and after privacy protection are very similar from various perspectives, ensuring the effectiveness of the primary BCI tasks.

\begin{figure}[!t]\centering
\subfigure[]{\label{fig:fig2a}  \includegraphics[width=0.97\linewidth,clip]{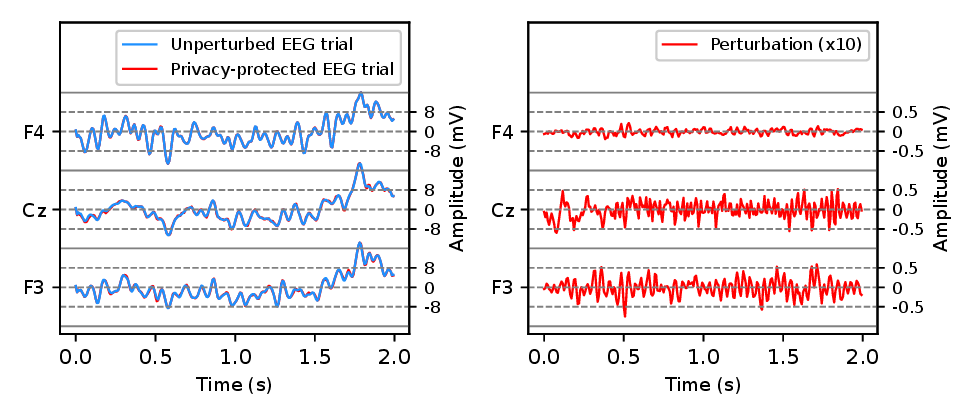}}
\subfigure[]{\label{fig:fig2b}  \includegraphics[width=0.97\linewidth,clip]{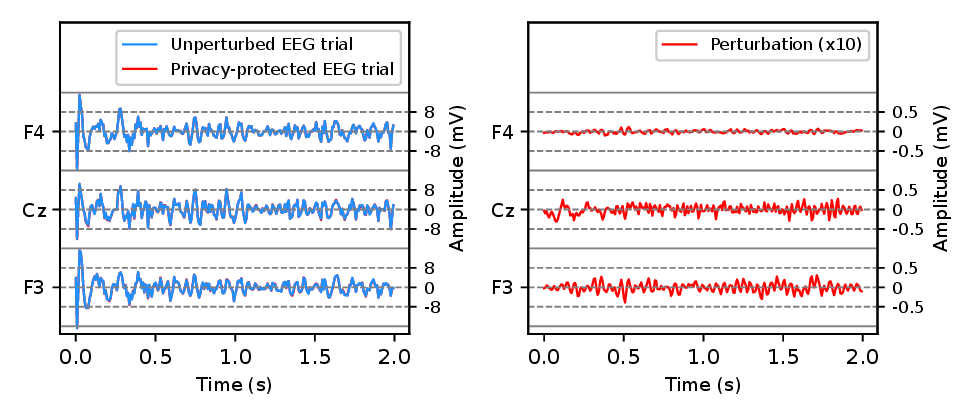}}
\subfigure[]{\label{fig:fig2c}  \includegraphics[width=0.97\linewidth,clip]{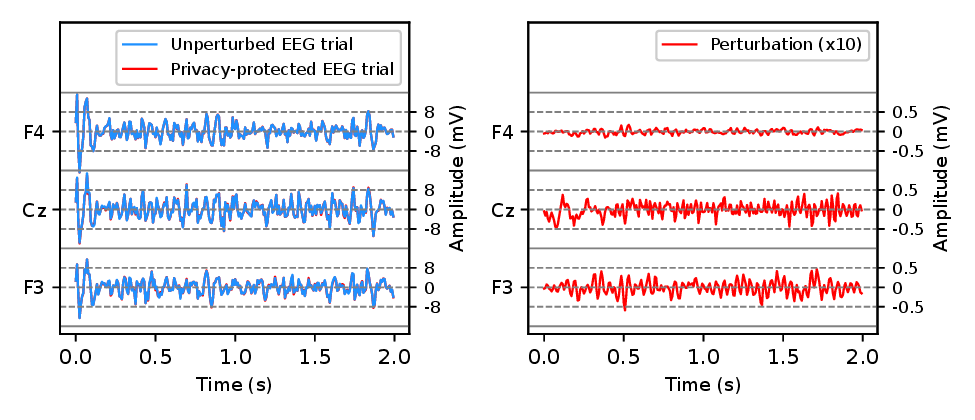}}
\caption{Original unperturbed EEG trials and their perturbed counterparts from the three BCI tasks, which are almost identical. (a) ERP; (b) MI; and, (c) SSVEP. The perturbations are magnified $10$ times for better visualization. The interval between each dashed line represents the amplitude of 8mV for EEG signals, and 0.5mV for perturbations.} \label{fig:2}
\end{figure}

Fig.~\ref{fig:2} shows the original unperturbed EEG trials and their perturbed counterparts from ERP, MI and SSVEP tasks. For clarity, only three EEG channels (F4, Cz, and F3) are shown. We can observe that the privacy-protected EEG trials and the original unperturbed EEG trials are almost identical for all three BCI tasks.

\begin{figure}[!t]\centering
\subfigure[]{\includegraphics[width=0.97\linewidth,clip]{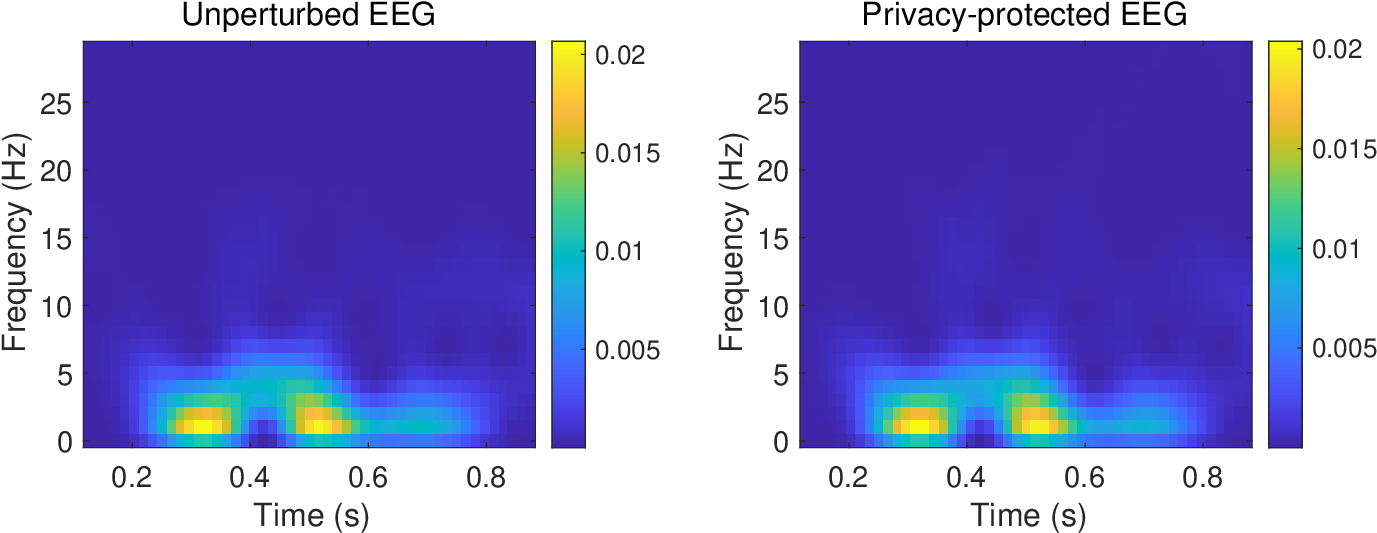}}
\subfigure[]{\includegraphics[width=0.97\linewidth,clip]{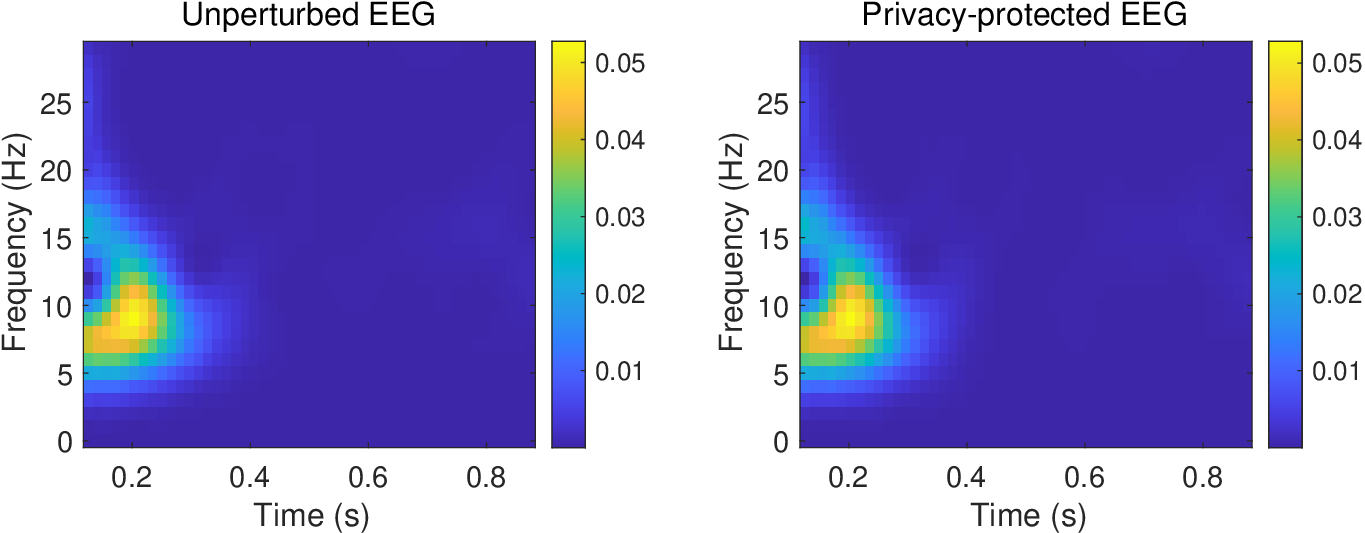}}
\subfigure[]{\includegraphics[width=0.97\linewidth,clip]{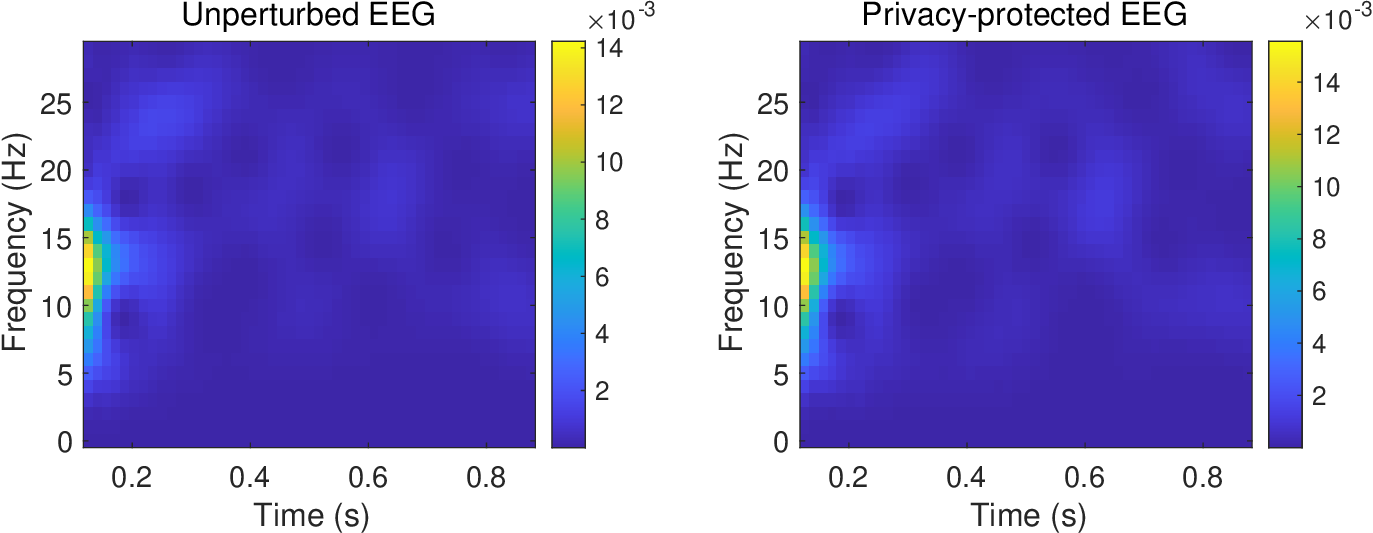}}
\caption{Average Cz channel spectrograms of the original unperturbed EEG trials and the privacy-protected EEG trials on three BCI tasks. (a) ERP; (b) MI; and, (c) SSVEP.} \label{fig:3}
\end{figure}

Fig.~\ref{fig:3} shows the average Cz channel spectrograms of the original unperturbed EEG trials and the privacy-protected EEG trials for the target class on the ERP task, the imagination of the right hand on the MI task, and the target with 12Hz flickering frequency on the SSVEP task. The perturbed spectrograms were almost identical to the unperturbed counterparts, suggesting that the perturbations hardly affect the spectrogram of the EEG trials.

\begin{figure}[htbp]\centering
\subfigure[]{\includegraphics[width=0.92\linewidth,clip]{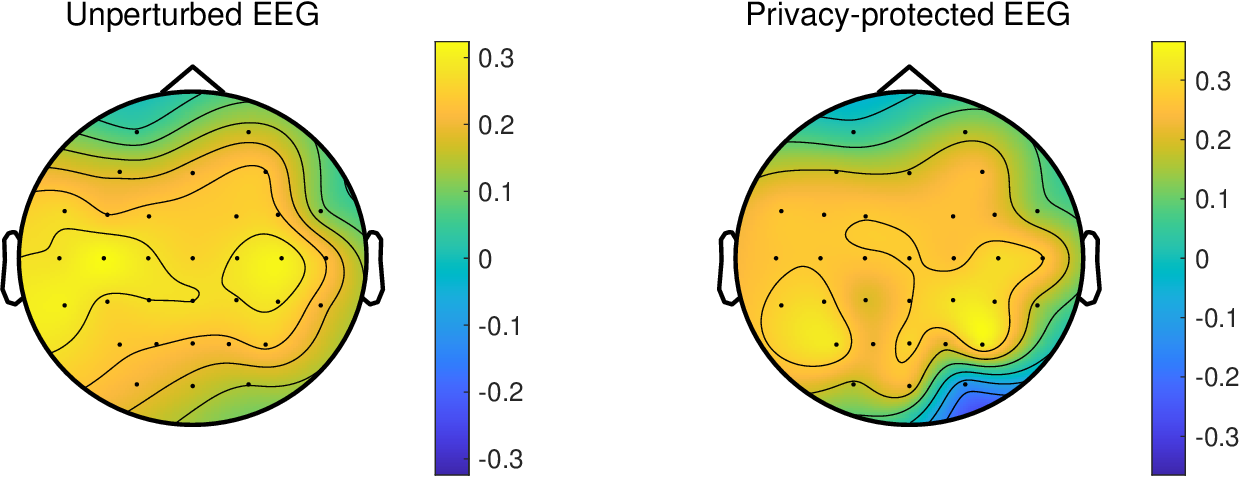}}
\subfigure[]{\includegraphics[width=0.92\linewidth,clip]{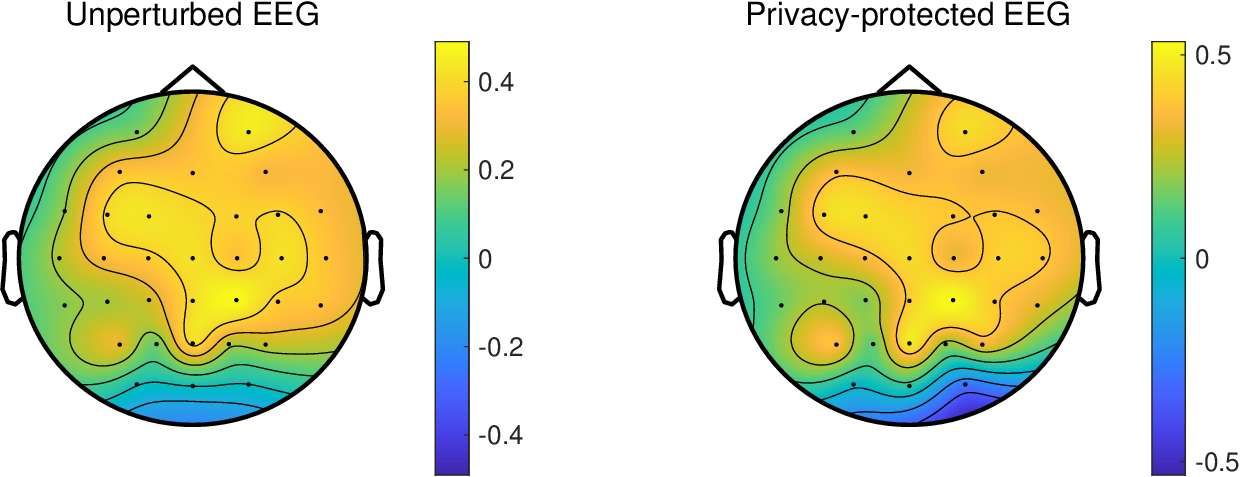}}
\subfigure[]{\includegraphics[width=0.92\linewidth,clip]{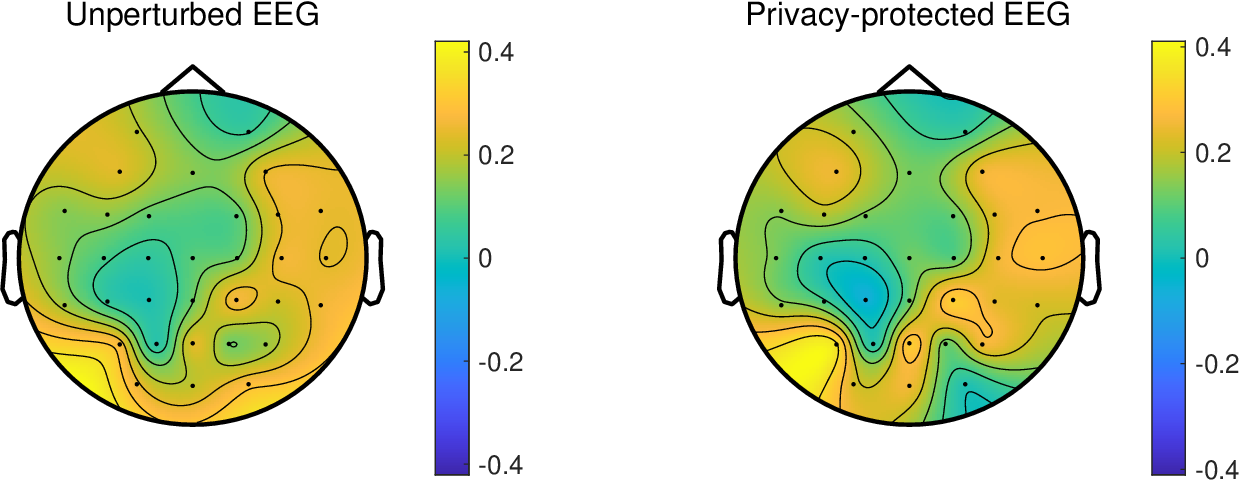}}
\caption{Average topoplots of the original unperturbed EEG trials and the privacy-protected EEG trials on three BCI tasks. (a) ERP; (b) MI; and, (c) SSVEP.} \label{fig:4}
\end{figure}

Fig.~\ref{fig:4} shows the average topoplots of the original unperturbed EEG trials and the privacy-protected EEG trials for the target class on the ERP task, the imagination of the right hand on the MI task, and the target with 12Hz flickering frequency on the SSVEP task. One can hardly observe any differences between the unperturbed and perturbed topoplots regardless of the BCI task, indicating that perturbations hardly change the spatial information of the EEG trials.

\subsection{Visualization of the Training Process}

\begin{figure}[htbp]\centering
\subfigure[]{\label{fig:5a}  \includegraphics[width=0.458\linewidth,clip]{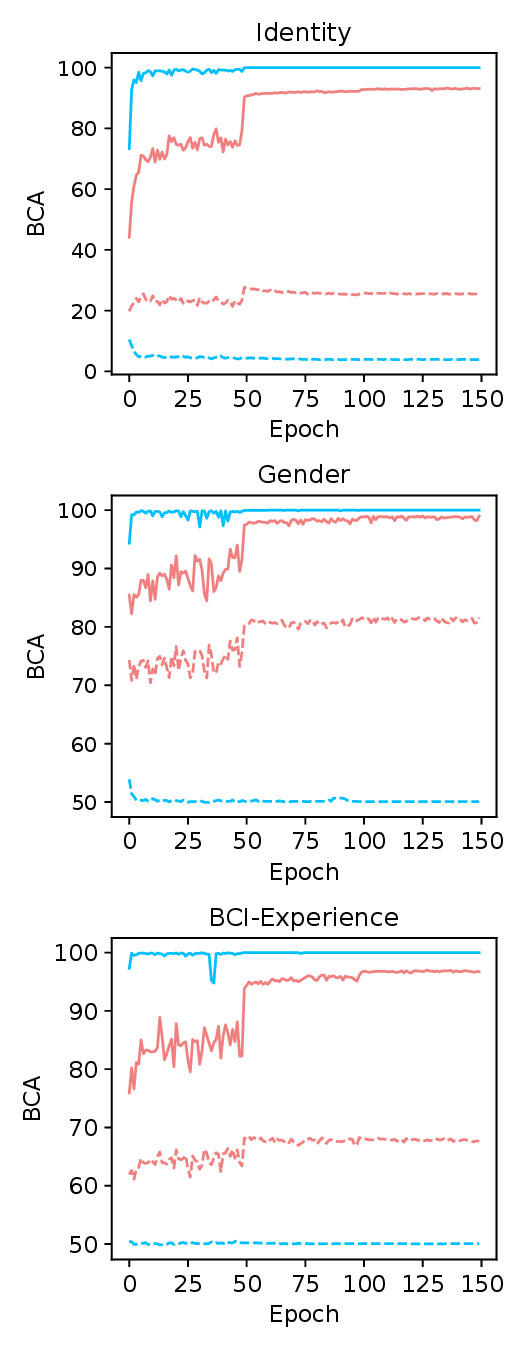}}
\subfigure[]{\label{fig:5b}  \includegraphics[width=0.497\linewidth,clip]{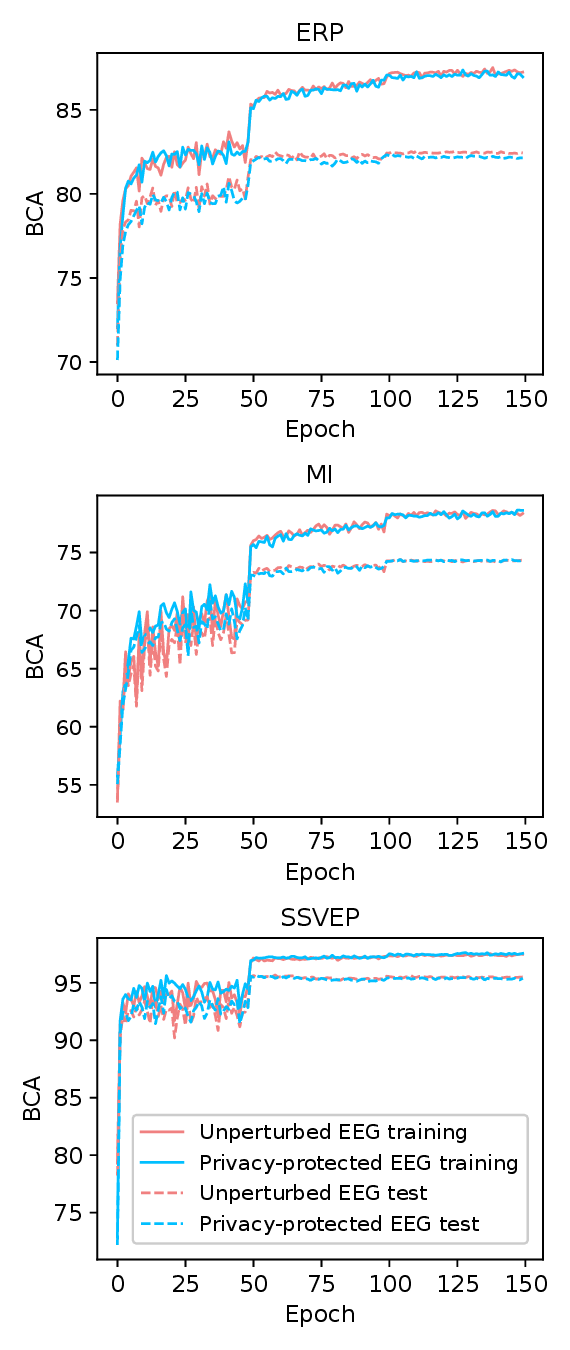}}
\caption{The training and test BCAs of three Privacy-Classifiers and three Task-Classifiers on the original unperturbed EEG data and the privacy-protected EEG data. (a) Privacy-Classifiers; and, (b) Task-Classifiers.} \label{fig:5}
\end{figure}

Fig.~\ref{fig:5} shows how the training and test BCAs of the Privacy-Classifiers and the Task-Classifiers change with the number of training epochs on the original unperturbed EEG data and their perturbed counterparts, respectively. Observe that:
\begin{enumerate}
	\item The training BCAs of the Privacy-Classifiers on the original unperturbed EEG data and the privacy-protected EEG data were close after convergence; however, the test BCAs of the Privacy-Classifiers trained on the privacy-protected EEG data were significantly lower than those trained on their unperturbed counterparts, suggesting that the private information in the EEG data had been successfully concealed.
	\item The BCA curves for both training and test of the Task-Classifiers on the original unperturbed EEG data and their perturbed counterparts nearly indistinguishable, indicating once again that the perturbations had little impact on the primary BCI tasks.
\end{enumerate}

\section{Conclusions}

There is rich private information in EEG signals, such as user identity, gender and BCI-experience, necessitating privacy protection in EEG-based BCIs. This paper has demonstrated that user's identity, gender, and BCI-experience can be easily inferred by machine learning models from EEG signals, exposing a serious privacy issue in EEG-based BCIs, which may significantly impact users' willingness to share their EEG data. To address this issue, we designed perturbations to convert the original EEG data into privacy-protected EEG data, which can conceal user private information while preserving the performance of the primary BCI tasks. Experimental results showed that the generated privacy-protected EEG data can significantly reduce simultaneously the classification accuracies on user identity, gender, and BCI-experience, while almost do not impact the primary BCI task classification performance at all.

%\section*{Acknowledgement}
%This research was supported by the Fundamental Research Funds for the Central Universities 2023BR024.

%\bibliographystyle{IEEEtran}\bibliography{Meng}


\end{document}